\documentclass[aps, prd, twocolumn, 
lengthcheck, superscriptaddress, 
nofootinbib]{revtex4-1}

\usepackage[normalem]{ulem}
\usepackage{color}
\usepackage{amsfonts}
\usepackage{amsmath}
\usepackage{amssymb}

\definecolor{ar}{rgb}{1.0, 0.01, 0.24}
\definecolor{al}{rgb}{0.82, 0.1, 0.26}
\definecolor{ev}{rgb}{0.56, 0.0, 1.0}
\def\be{\begin{eqnarray}}\def\ee{\end{eqnarray}}

\newcommand{\skl}[1]{{ #1 }}

\usepackage{amsmath,amssymb,amsfonts}
\usepackage{graphicx}
\usepackage{array}
\usepackage{longtable}
\usepackage{multirow}
\usepackage{booktabs}
\usepackage{bm}         
\usepackage{hyperref}   
\usepackage{physics}    

\graphicspath{{figures/}{../figures/}}

\begin{document}


\title{ NNStar: An end-to-end AI agent for nuclear matter and neutron star physics}

\author{Yao Ma}
\email{mayao@nju.edu.cn}
\affiliation{School of Frontier Sciences, Nanjing University, Suzhou 215163, China}

\author{Yong-Liang Ma}
\email{ylma@nju.edu.cn}
\affiliation{School of Physics, Nanjing University, Nanjing 210093, China}
\affiliation{School of Frontier Sciences, Nanjing University, Suzhou 215163, China}

\author{Jia-Ying Xiong}
\email{xiongjiaying21@mails.ucas.ac.cn}
\affiliation{School of Physics, Nanjing University, Nanjing 210093, China}
\affiliation{School of Frontier Sciences, Nanjing University, Suzhou 215163, China}

\date{\today}

\begin{abstract}

Constraining the equation of state of dense matter requires confronting effective models with massive data that spans many orders of magnitude in scale, from sub-saturation nuclear matter properties to the masses, radii, and tidal deformabilities of neutron stars.
Exploring the high-dimensional coupling space of such a model and fine tuning it against all of these constraints is a labor- and time-intensive task.
We present \textsc{NNStar}, an end-to-end artificial-intelligence agent that automates this workflow.
Rather than a bespoke application, \textsc{NNStar} is delivered as a portable \emph{skill} for an open large-language-model (LLM) agent platform---a self-describing module that pairs worked usage conventions with symbolic and numerical physics engines that (i) build a relativistic mean-field model directly from a Lagrangian, (ii) solve the mean-field equations of motion and evaluate the saturation properties, (iii) construct the $\beta$-equilibrium equation of state, splice it to a crust, and integrate the Tolman--Oppenheimer--Volkoff equations, and (iv) score the resulting predictions through a Bayesian joint analysis against nuclear matter and astrophysical observations.
The agent can read a model, fit its parameters, and report the full set of nuclear matter and neutron star observables without human intervention. \textsc{NNStar} therefore provides a new, AI-driven framework for analyzing nuclear matter and neutron-star observations.

\end{abstract}

\maketitle

\allowdisplaybreaks


\section{Introduction}
\label{sec:intro}

The properties of dense nuclear matter (NM), especially those relevant to the cores of massive neutron stars (NSs), have been studied for several decades. Nevertheless, several fundamental questions remain open, such as the composition of the matter and the evolution of the symmetries of QCD~\cite{Fukushima:2013rx,Holt:2014hma,Baym:2017whm,Ma:2019ery,Burgio:2021vgk,Ma:2021nuf,Brandes:2023bob,Gupta:2026vov}.
The observation of massive NSs around $2M_\odot$~\cite{Demorest:2010bx,Antoniadis:2013pzd,Fonseca:2021wxt}, the detection of gravitational waves from the binary NS merger GW170817~\cite{LIGOScientific:2017vwq,LIGOScientific:2018cki}, and the simultaneous mass--radius (M-R) measurements by NICER~\cite{Vinciguerra:2023qxq,Salmi:2024aum,Choudhury:2024xbk,Mauviard:2025dmd} have together sharpened the constraints on the equation of state (EoS) while simultaneously introducing tensions that are increasingly difficult to reconcile within a single theoretical framework.
On the terrestrial side, heavy-ion collisions (HICs) and the analysis of nuclear structure provide complementary information on NM properties around the saturation density $n_0 \approx 0.16~\mathrm{fm}^{-3}$~\cite{Dutra:2014qga,lattimer2013constraining,Danielewicz:2013upa,LeFevre:2015paj,Brown:2013mga,Reed:2021nqk}.

To confront these data, the EoS is usually parameterized through an effective model or theory whose low-energy constants encode the underlying QCD dynamics and are fixed by fitting the data.
The fitting process, however, is delicate. Because, for example, a fine tuning of the attractive and repulsive forces is needed to pin down the small binding energy and the location of the saturation density, the physical observables depend sensitively on the couplings, and a global agreement with data across broad scales can only be reached by a careful, and often tedious, exploration of a high-dimensional parameter space. The simultaneous treatment of massive heterogeneous data and the fine tuning of the couplings are therefore very difficult, if not unfeasible, to accomplish by hand.

Machine-learning (ML) and artificial-intelligence (AI)-driven techniques offer a way out, owing to their efficiency and adaptivity in analyzing massive data and uncovering statistical correlations~\cite{vaswani2017attention,Radovic:2018dip,Carleo:2019ptp,goodfellow2020generative,Boehnlein:2021eym,Zhou:2023pti}. In nuclear astrophysics, they have already been used extensively to constrain the NM EoS and NS structure through deep neural networks (NNs), conditional variational autoencoders, normalizing flows, and Bayesian deep learning that map astrophysical observables to the interior EoS or directly to the speed of sound~\cite{Fujimoto:2019hxv,Wang:2020tgb,Morawski:2020izm,Anil:2020lch,Wu:2021tri,Han:2021kjx,Fujimoto:2022ohj,Soma:2022qnv,OmanaKuttan:2022aml,He:2023urp,Guo:2023mhf,Zhou:2023pti,Chatterjee:2023ecc,He:2023zin,Guo:2024nzi,Yuksel:2024zky,Ventagli:2024xsh,Ferreira:2024rnf,Zhou:2024yzy,Patra:2025xtd,Patra:2025xtd,Kovlakas:2025igr,Liu:2026dxn,Baker:2026phm,Bezerra:2026eun}. These efforts span several methodological families. Supervised deep networks learn the inverse map from M-R and tidal-deformability data to the EoS or to NM parameters~\cite{Fujimoto:2019hxv,Morawski:2020izm}; Bayesian and nonparametric neural representations relax the assumed functional form of $P(\varepsilon)$ while propagating observational and epistemic uncertainties~\cite{Han:2021kjx,Ventagli:2024xsh}; generative models such as conditional variational autoencoders enable rapid, calibrated posterior sampling of EoSs compatible with a given set of observations~\cite{Ferreira:2024rnf}; and differentiable or symbolic surrogates connect the microscopic EoS or the sound speed $c_s^2(n)$ to macroscopic stellar observables and accelerate the forward Tolman--Oppenheimer--Volkoff (TOV) map~\cite{Soma:2022qnv,Chatterjee:2023ecc,Patra:2025xtd}. These studies each learn or emulate an individual statistical or physical mapping, rather than autonomously executing the full physics pipeline from a chosen Lagrangian to the final constraint. In our earlier work~\cite{Guo:2023mhf}, we constructed a modular NN platform that, taking a Walecka-type relativistic mean-field (RMF) model as a concrete example, estimates the model parameters such that both the NM properties around $n_0$ and the global NS properties are reproduced, and showed that the platform extends straightforwardly to other effective models. This work demonstrated the feasibility of automating the parameter search. What remained was to build an autonomous agent that itself plans and drives the entire physics pipeline---from model construction to the final M-R curve and its statistical assessment---so that the strong interaction in dense matter can be investigated end-to-end with minimal human steering.

Building such an autonomous agent---one that plans and drives the entire physics pipeline---is timely. The large-language-model (LLM) advances that power general-purpose assistants have given rise to autonomous scientific agents that reason, call external tools, and execute multi-step research workflows~\cite{Yao:2022react,Schick:2023tool,Zheng:2025survey}, with demonstrations ranging from autonomous chemical synthesis~\cite{Bran:2023chemcrow,Boiko:2023nature} to end-to-end machine-learning research~\cite{Lu:2024aiscientist,Yamada:2025aiscientist2}.
A recurring lesson from these efforts is that the reliability of an agent hinges on grounding: replacing free-form recall with calls to verifiable, deterministic tools.

\begin{figure}[t]
\centering
\includegraphics[width=0.48\textwidth]{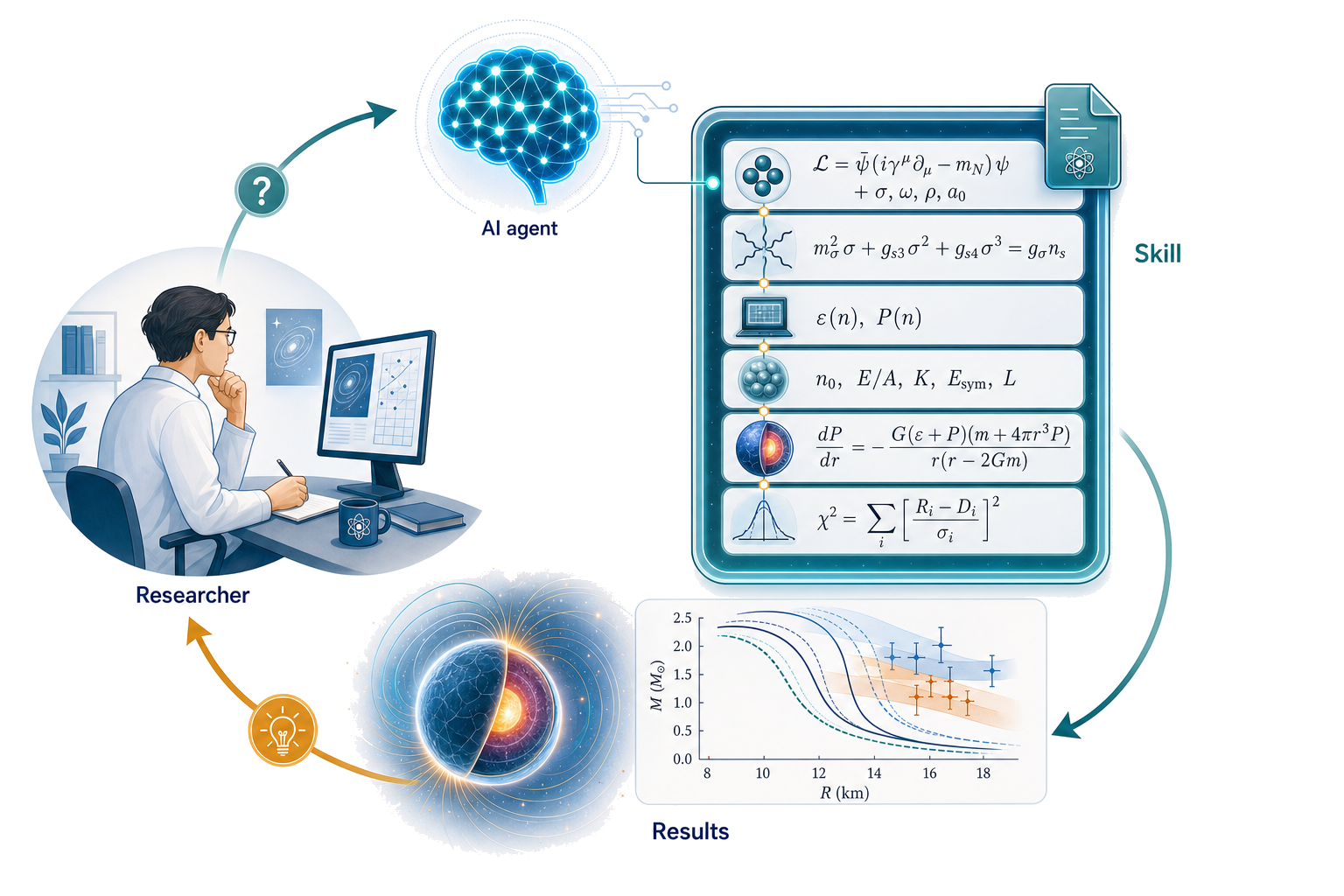}
\caption{The \textsc{NNStar} workflow as a closed research loop. A researcher poses a natural-language question to an LLM agent, which drives the bundled \textsc{NNStar} skill---a documented, model-agnostic library that builds a RMF model from a Lagrangian, solves the mean-field EoM, constructs the EoS, evaluates the saturation properties, integrates the Tolman--Oppenheimer--Volkoff equations to the M-R curve, and scores the result through the Bayesian joint analysis. The fitted couplings, NM properties, and M-R curves are returned to the researcher, closing the loop.}
\label{fig:workflow}
\end{figure}

In this work, we take that step---constructing an LLM agent that autonomously plans and executes the full forward-and-inference pipeline, from a constructed model to the final M-R curve and its statistical assessment---and introduce \textsc{NNStar}, an end-to-end AI agent for NM and NS physics, realized not as a bespoke program but as a portable \emph{skill} that any LLM agent can load. The design rests on four guiding principles, illustrated schematically in Fig.~\ref{fig:workflow}:
(i) a single model---the agent operates on one self-consistent Lagrangian at a time, building all observables from first principles rather than from disconnected fits;
(ii) fine tuning of the equations of motion (EOMs)---the agent owns the symbolic derivation of the mean-field EOMs and the numerical solver, so that the couplings are tuned against the exact field equations of the chosen model; (iii) extensibility---new models, constraints, and tools can be added without rewriting the agent, because the physics capabilities are exposed as discrete tools and the models are described by a generic, declarative format; and
(iv) portability---the entire capability is delivered as a single, platform-agnostic skill (a documented specification bundled with executable physics code), so that any compliant LLM agent acquires it by loading the skill rather than through bespoke integration.

The rest of this paper places \textsc{NNStar} in the context of autonomous scientific agents and ML for dense matter (Sec.~\ref{sec:related}), describes the model (Sec.~\ref{sec:model}), the architecture of the agent and the tools it controls (Sec.~\ref{sec:agent}), and the computational physics methods implemented in the skill (Sec.~\ref{sec:tools}).
We then exercise the skill along three steps that illustrate how it would be used in practice: we validate it by reproducing established Walecka-type models, demonstrate its extensibility by adding a sextic scalar self-interaction to TM1 and re-optimizing the enlarged model against data (Sec.~\ref{sec:validation}), and present a compact benchmark of several contemporary LLMs driving the same skill (Sec.~\ref{sec:benchmark}). The implications of these validation, application, and benchmark results for the use of grounded, skill-based agents in NM and NS studies are discussed in Sec.~\ref{sec:conclusion}.

\section{Background and related works}
\label{sec:related}

\textsc{NNStar} sits at the intersection of three lines of research: tool-augmented LLMs, autonomous scientific agents, and ML for the dense-matter EoS. We summarize each line and clarify what \textsc{NNStar} adds in this section.

{\bf Tool-augmented LLMs.}---A central insight of recent LLM research is that reasoning quality improves dramatically when the model can interleave thought with actions that query external tools.
The ReAct paradigm couples chain-of-thought reasoning with tool calls~\cite{Yao:2022react}, while Toolformer shows that models can learn when and how to invoke APIs~\cite{Schick:2023tool}.
Standardizing this interface, the Model Context Protocol (MCP) defines a client--server convention in which tool providers expose ``tools'', ``resources'', and ``prompts'' that any compliant LLM can discover and call~\cite{Anthropic:2024mcp}. Building on this interface, recent agent platforms expose reusable \emph{skills}---self-contained folders that bundle a documented specification with executable code, which the platform surfaces to the model on demand~\cite{Anthropic:2025skills}.
\textsc{NNStar} packages its entire physics backend as such a skill, so that the backend is decoupled from, and interchangeable with the reasoning model, and portable across any skill-aware agent platform.

{\bf Autonomous scientific agents.}---Building on tool use, several systems now carry out multi-step research with limited human input.
In chemistry, ChemCrow equips an LLM with expert tools for synthesis planning and execution~\cite{Bran:2023chemcrow}, and a related platform autonomously designs and runs reactions~\cite{Boiko:2023nature}.
In ML research itself, ``The AI Scientist'' and its successor automate hypothesis generation, experimentation, and manuscript writing~\cite{Lu:2024aiscientist,Yamada:2025aiscientist2}; broad surveys chart the move from automation to autonomy and catalogue the recurring failure modes---bias toward training defaults, implementation drift, and weak verification~\cite{Zheng:2025survey}.
Closer to the physical sciences, autonomous agents have begun to target computation and experiment directly, from quantum-chemistry agents that generate and run workflows from natural-language prompts~\cite{Zou:2025elagente} to self-driving laboratories that plan and execute the synthesis of novel inorganic materials~\cite{Szymanski:2023alab}.
Compared with these largely domain-general or chemistry-focused systems, \textsc{NNStar} is a domain-specific skill whose tools encode the exact field equations of a chosen Lagrangian, so that the outputs of any agent that loads it are reproducible and physically auditable rather than merely plausible.

{\bf ML for the dense-matter EoS.}---Within nuclear astrophysics, ML has been used to infer the EoS from NS observables with deep networks~\cite{Ventagli:2024xsh,Patra:2025xtd}, conditional variational autoencoders~\cite{Ferreira:2024rnf}, and Bayesian deep learning for the speed of sound~\cite{Fujimoto:2022ohj,OmanaKuttan:2022aml}, complementing the NN parameter-search platform of Ref.~\cite{Guo:2023mhf}. These works learn a statistical map between observables and the EoS. \textsc{NNStar} is complementary: rather than learning the map, it automates the forward modeling and inference pipeline for a physically transparent Lagrangian, with the LLM supplying the planning and the deterministic tools supplying the physics.

\section{The model}
\label{sec:model}

To set up the physics backbone of \textsc{NNStar}, we consider the general quantum hadrodynamical (GQHD) model that incorporates all hadron resonances below $1.0$~GeV, i.e., the $\sigma$, $\omega$, $\rho$, and $a_0$ mesons in addition to the nucleons~\cite{Ma:2026kun}. Note that this choice fixes only the adopted \emph{base model}, not the framework itself. The underlying skill is model-agnostic, deriving and solving the field equations of whatever Lagrangian it is supplied. The GQHD model is used here because it is general enough to contain the standard RMF parameterizations as special cases (see below). The pion is not included because it vanishes in the RMF approximation. Keeping all operators with (iso-)scalars and vectors up to dimension-four, the Lagrangian reads
\be
\mathcal{L} & = & \mathcal{L}_N+\mathcal{L}_\sigma+\mathcal{L}_\omega+\mathcal{L}_\rho+\mathcal{L}_{a_0}+\mathcal{L}_I,
\ee
where
\be
\mathcal{L}_N & = & \bar{\Psi}\left(i \gamma^\mu \partial_\mu-m_N\right) \Psi, \nonumber\\
\mathcal{L}_\sigma & = & \frac{1}{2}\left(\partial_\mu \sigma \partial^\mu \sigma-m_\sigma^2 \sigma^2\right)-\frac{1}{3} g_{s3} \sigma^3-\frac{1}{4} g_{s4} \sigma^4 \nonumber\\
& &{} -\frac{1}{2} g_{s2o2} \sigma^2 \omega^\mu \omega_\mu-\frac{1}{2} g_{s2r2} \sigma^2 \vec{\rho}^\mu \cdot \vec{\rho}_\mu \nonumber\\
& &{} -g_{sr2} \sigma \vec{\rho}^\mu \cdot \vec{\rho}_\mu-g_{so2} \sigma \omega^\mu \omega_\mu, \nonumber\\
\mathcal{L}_\omega & = & -\frac{1}{4} \Omega^{\mu \nu} \Omega_{\mu \nu}+\frac{1}{2} m_\omega^2 \omega^\mu \omega_\mu+\frac{1}{4} g_{o4}\left(\omega^\mu \omega_\mu\right)^2 \nonumber\\
& &{} +\frac{1}{2} g_{o2r2} \omega^\mu \omega_\mu \vec{\rho}^\mu \cdot \vec{\rho}_\mu, \nonumber\\
\mathcal{L}_\rho & = & -\frac{1}{4} \vec{P}^{\mu \nu} \cdot \vec{P}_{\mu \nu}+\frac{1}{2} m_\rho^2 \vec{\rho}^\mu \cdot \vec{\rho}_\mu+\frac{1}{4} g_{r4}\left(\vec{\rho}^\mu \cdot \vec{\rho}_\mu\right)^2, \nonumber\\
\mathcal{L}_{a_0} & = & \frac{1}{2} \partial_\mu \vec{a}_0 \cdot \partial^\mu \vec{a}_0-m_{a_0}^2 \vec{a}_0 \cdot \vec{a}_0+\frac{1}{4} g_{a4}\left(\vec{a}_0 \cdot \vec{a}_0\right)^2 \nonumber\\
& &{} +\frac{1}{2} g_{a2r2}\left(\vec{a}_0 \cdot \vec{a}_0\right)\left(\vec{\rho}^\mu \cdot \vec{\rho}_\mu\right)+\frac{1}{2} g_{sa2} \sigma\left(\vec{a}_0 \cdot \vec{a}_0\right) \nonumber\\
& &{} +\frac{1}{2} g_{s2a2} \vec{a}_0 \cdot \vec{a}_0 \sigma^2+\frac{1}{2} g_{o2a2} \vec{a}_0 \cdot \vec{a}_0 \omega^\mu \omega_\mu \nonumber\\
& &{} + g_{sora} \vec{a}_0 \cdot \vec{\rho}^\mu \omega_\mu \sigma+g_{ora} \vec{a}_0 \cdot \vec{\rho}_\mu \omega^\mu, \nonumber\\
\mathcal{L}_I &= & \bar{\Psi}\left(g_\sigma \sigma-g_\omega \gamma^\mu \omega_\mu-g_\rho \gamma^\mu \vec{\rho}_\mu+g_{a_0} \vec{a}_0\right) \Psi ,
\label{eq:lagrangian}
\ee
where $\Psi=(p,n)^T$ is the iso-doublet nucleon field, $\vec{\rho}_\mu=\rho^i_\mu\tau^i$ and $\vec{a}_0=a^i_0\tau^i$, and $\Omega^{\mu\nu}$ and $\vec{P}^{\mu\nu}$ are the field-strength tensors of the vector mesons. Together with the masses fixed to their physical values, the free couplings span a $21$-dimensional parameter set $\theta$~\cite{Ma:2026kun}. This Lagrangian is general enough to contain many Walecka-type models, such as TM1~\cite{Sugahara:1993wz}, NL3~\cite{Lalazissis:1996rd}, and FSU~\cite{Todd-Rutel:2005yzo} as their special cases, which makes it the natural common substrate for the agent. Every model the agent handles is a particular truncation of, or a particular point in, the coupling space of Eq.~\eqref{eq:lagrangian}.

A defining feature of the GQHD model is the mixed $\sigma\omega\rho a_0$ coupling.
As shown in Ref.~\cite{Ma:2026kun}, this single interaction substantially reshapes the speed of sound and the M-R relation under charge neutrality and $\beta$-equilibrium, producing a peak in the sound speed at $\sim(2$--$3)\,n_0$. 
Such a non-monotonic sound speed---rising well above the conformal value $v_s^2=1/3$ at intermediate densities before approaching it from above at higher densities---is required by the joint existence of $2M_\odot$ stars and small intermediate-mass radii. It has been widely regarded as a signature of the transitions from hadrons to other compositions such as quark and quarkyonic matter~\cite{Bedaque:2014sqa,McLerran:2018hbz,Tan:2020ics,Altiparmak:2022bke}. That the GQHD model reproduces this feature within a purely hadronic description is therefore quite interesting.

\section{The \textsc{NNStar} agent}
\label{sec:agent}

\textsc{NNStar} is structured as a thin reasoning layer---the LLM---on top of a deterministic physics stack. The LLM never computes physics itself; instead, it plans a sequence of tool calls, inspects structured results, and decides on subsequent steps. This separation ensures that numerical outputs remain reproducible and auditable, while allowing the agent to handle open-ended tasks---such as choosing models, selecting constraints, interpreting failures---in natural language.

\subsection{Architecture}

\textsc{NNStar} is not a standalone program, but rather a skill installed into an open agent platform (here, \texttt{openclaw}). The platform runs the agentic loop against the LLM and advertises the installed skills to the model. When a relevant request is received, the model reads the skill's specification (its \texttt{SKILL.md}) and invokes the bundled physics code through the platform's code-execution tool. Because the capability travels with the skill rather than with a bespoke backend, the same skill remains portable across any skill-aware platform and any sufficiently capable LLM. Moreover, extending or swapping the physics requires only edits the skill folder. This design also aligns \textsc{NNStar} with the emerging tool-use and agent-skill ecosystem~\cite{Yao:2022react,Schick:2023tool}. Schematically,
\begin{equation*}
{\begin{aligned}
&\text{User} \;\rightarrow\; \text{agent platform} \;\rightleftarrows\; \text{LLM},\\
&\text{LLM} \;\rightarrow\; \text{skill}~(\texttt{SKILL.md}+\text{code}) \;\rightarrow\; \text{execution}.
\end{aligned}}
\end{equation*}
This design makes the skill model-agnostic on both ends: the LLM can be replaced (Sec.~\ref{sec:benchmark}) and the physics library can be extended independently.

\subsection{Physics tool layers}

Beneath the skill's documented interface lies a layered, model-agnostic computation framework:
\begin{itemize}
\item {\bf Analytic layer.}---Given the parameters, fields, and Lagrangians (as symbolic expressions), this layer automatically constructs the mean-field EOMs, the effective masses, and the energy-density terms using a computer-algebra engine. Because the EOMs are derived symbolically from the input Lagrangian, the same code can handle any model expressible in the form of Eq.~\eqref{eq:lagrangian}---or more general forms---with the same framework.
\item {\bf Numeric layer.}---For a given baryon density, this layer solves the coupled mean-field equations for the meson fields, evaluates the energy density, and returns the field solution, effective masses, and scalar densities.
\item {\bf Nuclear-properties layer.}---From the energy density $\varepsilon(n_n,n_p)$ with $n_n$ and $n_p$ being, respectively, the neutron and proton densities, this layer computes the saturation observables: the saturation density $n_0$, binding energy $E$, incompressibility $K$, symmetry energy $E_{\rm sym}$, and its slope $L$.
\item {\bf TOV layer.}---This layer solves for $\beta$-equilibrium matter, splices the high-density EoS onto the low-density BPS crust to form a hybrid EoS, and integrates the TOV equations to produce the M-R curve.
\end{itemize}
Models are described declaratively in a JSON format (including parameter keys, field names, Lagrangian term strings, parameter values, and the density grid), and a generic loader transforms such a definition into a fully built model.
A single driver then runs the entire pipeline---nuclear properties $\rightarrow$ $\beta$-equilibrium EoS $\rightarrow$ crust-core (spliced) EoS$\rightarrow$ M-R curve
---for any model definition.
This forward chain, from a Lagrangian to the M-R curve, is the deterministic core that the skill bundles.
The higher-level actions described next---fitting, scanning, statistical assessment, and reading a model---are not separate engines, but rather workflows that the agent composes by repeatedly driving this core.

\subsection{What the skill lets the agent do}

Guided by the skill's \texttt{SKILL.md}, the agent drives the bundled forward engine to execute the workflow's high-level tasks---invoking it for a single evaluation, or wrapping it in a loop for optimization and exploration. Broadly, the skill lets an agent:
\begin{itemize}
\item {\bf Define or import a model}---assemble a model from a Lagrangian and its couplings (supplied directly, drawn from the built-in parameterizations, or reconstructed from a published paper, e.g., a PDF file).
\item {\bf Compute the observables}---run the forward pipeline end-to-end to obtain the NM properties, the $\beta$-equilibrium EoS, and the M-R curve for any model.
\item {\bf Tune and explore}---optimize the couplings against a chosen set of constraints, and scan individual couplings to map the response of the observables.
\item {\bf Assess and report}---score a model statistically against empirical and astrophysical data, and save, retrieve, and present the results.
\end{itemize}
The ability to import a model directly from the literature closes the loop between the published work and computation: the agent can be given a paper, reconstruct the model defined therein, and independently reproduce its NM and NS predictions.
This capability underpins the validation presented in Sec.~\ref{sec:validation}.

\section{Physics methods behind the skill}
\label{sec:tools}

\subsection{Mean-field approach}
\label{sec:mft}

The default solver implements the RMF approximation~\cite{Walecka:1974qa,Serot:1984ey}. In the simplest $\sigma$-$\omega$ model, the nucleon Dirac equation in uniform matter reads
\begin{equation}
\left(i\gamma^\mu\partial_\mu - g_\omega\gamma^0\omega_0 - m_N^*\right)\psi = 0, \quad
m_N^* = m_N - g_\sigma\phi,
\end{equation}
with the meson mean fields determined self-consistently by
\begin{equation}
\begin{aligned}
\phi &= \frac{g_\sigma}{m_\sigma^2}\,n_s,\quad n_s=\langle\bar\psi\psi\rangle, \\
\omega_0 &= \frac{g_\omega}{m_\omega^2}\,n_B,\quad n_B=\langle\psi^\dagger\psi\rangle .
\end{aligned}
\end{equation}
For model~\eqref{eq:lagrangian}, the analytic layer automatically generates the corresponding coupled equations. The current skill retains only the direct (Hartree) contributions, and its layered, model-agnostic design is built to accommodate the relativistic Hartree--Fock (RHF) method~\cite{Bouyssy:1987sh} as a drop-in extension, in which the exchange (Fock) contributions are retained and the nucleon self-energy acquires scalar, time, and space components, $\Sigma(p)=\Sigma_S(p)+\gamma^0\Sigma_0(p)+\bm{\gamma}\cdot\hat{\bm{p}}\,\Sigma_V(p)$. Once added, the Hartree and Hartree--Fock options together would allow the agent to quantify the role of exchange terms within the same model.

\subsection{TOV equation and the stellar model}
\label{sec:tov}

Given the $\beta$-equilibrium EoS, the NS structure follows from the TOV equation~\cite{Tolman:1939jz,Oppenheimer:1939ne}
\be
\frac{dP}{dr} & = &{} -\frac{G\mathcal{M}(r)\varepsilon}{r^2}\left(1+\frac{P}{\varepsilon}\right)\left(1+\frac{4\pi r^3 P}{\mathcal{M}(r)}\right) \nonumber\\
& & {} \times\left(1-\frac{2G\mathcal{M}(r)}{r}\right)^{-1}.
\ee
The star is modeled as a pure-hadronic-matter inner core, matched onto the BPS crust at low density~\cite{Baym:1971pw,Arnett:1977czg}.
Repeating the integration over a range of central pressures yields the full M-R curve, and the tidal deformability is obtained from the same background by integrating the perturbation equation.

\subsection{Bayesian joint analysis}
\label{sec:bja}

Taking the observables produced by the previous procedure as input, the statistical assessment of a model is performed with the Bayesian joint analysis (BJA) framework~\cite{Ma:2026kun,bayes1763essay}. By Bayes' theorem,
\begin{equation}
\begin{aligned}
p(\theta\mid D)&=\frac{p(D\mid\theta)\,p(\theta)}{p(D)}, \\
p(D)&=\int p(D\mid\theta)\,p(\theta)\,d\theta ,
\end{aligned}
\label{eq:bayes}
\end{equation}
where $\theta$ are the model parameters, $p(\theta)$ the prior (temporarily taken as uniform within the allowed ranges for simplicity), $p(D\mid\theta)$ the likelihood, and $p(D)$ the evidence. Assuming independent data, the likelihood factorizes
\begin{equation}
p(D\mid\theta)=\prod_{i=1}^{n} p(D_i\mid\theta) ,
\label{eq:like}
\end{equation}
due to the fact that the evidence $p(D)$ is hard to compute because of the high dimensionality of $\theta$.
The detailed normalization of the residuals and the definition of BJA score used to rank models are given in App.~\ref{app:BJA}.

\subsection{Phenomenological constraints}
\label{sec:constraints}

The data $D$ entering Eqs.~\eqref{eq:bayes}--\eqref{eq:like} consists of two complementary blocks, both imposed at the $1\sigma$ level:
\begin{itemize}
\item NM properties around $n_0$: the binding energy $E(n)$, pressure $P(n)$, incompressibility $K(n)$, symmetry energy $E_{\rm sym}(n)$, and its slope $L(n)$, taken from analyses of nuclear structure and HICs~\cite{Dutra:2014qga,Dutra:2012mb,lattimer2013constraining,Lattimer:2012xj,Danielewicz:2013upa,LeFevre:2015paj,Brown:2013mga,Reed:2021nqk};
\item NS M-R relations: the constraints from the massive pulsars and the NICER measurements PSR~J0030+0451, PSR~J0437--4715, and PSR~J0614--3329, as well as the GW170817 tidal deformability~\cite{Antoniadis:2013pzd,LIGOScientific:2017vwq,Vinciguerra:2023qxq,Choudhury:2024xbk,Mauviard:2025dmd,LIGOScientific:2018cki,Salmi:2024aum}.
\end{itemize}
Within the BJA, these two blocks are combined into a single likelihood, so that the couplings are constrained simultaneously and consistently by both NM data and astrophysical observations.
The agent exposes this constraint catalog through the \emph{list the data constraints} operation and lets the user select any subset for a given fit.

\section{Validation and application}
\label{sec:validation}

\subsection{Validation on Walecka-type models}
\label{sec:walecka-val}

As a first check, the agent is run on established Walecka-type models---TM1~\cite{Sugahara:1993wz}, NL3~\cite{Lalazissis:1996rd}, and the $\delta$-meson-extended FSU model FSU-$\delta$6.7~\cite{Todd-Rutel:2005yzo}---by extracting their definitions and recomputing both the NM properties and M-R relations.
Recovering the published saturation properties and M-R curves for these models confirms that the analytic derivation of the EOMs, the numerical solver, the crust matching, and the TOV integration are all correct end-to-end.

Table~\ref{tab:val-nmp} compares the saturation properties recovered by the skill with the values obtained in Ref.~\cite{Ma:2026kun}. The agent reproduces the saturation density $n_0$, binding energy $E$, symmetry energy $E_{\rm sym}$, and its slope $L$ to within the rounding of the published couplings (typically $\lesssim2\%$), and the incompressibility $K$ to within $\sim7\%$ (see below). Figure~\ref{fig:val-mr} overlays the corresponding agent-generated M-R curves on the current astrophysical constraints---the massive-pulsar mass of PSR~J0740+6620~\cite{Fonseca:2021wxt}, the NICER mass--radius measurements of PSR~J0030+0451~\cite{Vinciguerra:2023qxq}, PSR~J0740+6620~\cite{Salmi:2024aum}, and PSR~J0437$-$4715~\cite{Choudhury:2024xbk}, and the GW170817 tidal constraint on $R_{1.4}$~\cite{LIGOScientific:2018cki}. The agent recovers the well-known ordering of these forces: NL3 and TM1 are stiff and predict large radii ($R_{1.4}\gtrsim14.5$~km), in tension with the NICER and GW170817 radii, whereas the $\delta$-meson model FSU-$\delta$6.7 is softer, with $R_{1.4}\simeq13$~km and $M_{\rm max}\simeq2.09\,M_\odot$, passing close to the J0030 and J0740 ellipses---exactly the published behavior that the skill is meant to reproduce.

\begin{table*}[t]

\caption{\label{tab:val-nmp}Validation of the \textsc{NNStar} skill on three established Walecka-type models. For each model the saturation properties recovered by the agent (``NNStar'') are compared with the published values of Ref.~\cite{Ma:2026kun} (``Ref.''): the saturation density $n_0$, binding energy $E$, symmetry energy $E_{\rm sym}$, slope $L$, and incompressibility $K$. The neutron-star maximum mass $M_{\rm max}$ and the radius at $1.4\,M_\odot$, $R_{1.4}$, are the agent's predictions.}
\begin{ruledtabular}
\begin{tabular}{llccccc cc}
Model & Source & $n_0$ & $E$ & $E_{\rm sym}$ & $L$ & $K$ & $M_{\rm max}$ & $R_{1.4}$ \\
 & & [fm$^{-3}$] & [MeV] & [MeV] & [MeV] & [MeV] & [$M_\odot$] & [km] \\
\hline
TM1            & NNStar & $0.145$ & $-15.9$ & $36.6$ & $109.9$ & $275$ & $2.21$ & $14.60$ \\
               & Ref.   & $0.142$ & $-16.2$ & $36.0$ & $108$   & $257$ & --     & -- \\
\hline
NL3            & NNStar & $0.151$ & $-16.6$ & $38.7$ & $122.9$ & $264$ & $2.78$ & $14.85$ \\
               & Ref.   & $0.148$ & $-16.3$ & $38.2$ & $121$   & $246$ & --     & -- \\
\hline
FSU-$\delta$6.7 & NNStar & $0.149$ & $-16.4$ & $32.2$ & $52.5$  & $231$ & $2.09$ & $12.98$ \\
               & Ref.   & $0.148$ & $-16.3$ & $32.7$ & $53.5$  & $229$ & --     & -- \\
\end{tabular}
\end{ruledtabular}
\end{table*}

\begin{figure}[t]
\centering
\includegraphics[width=0.48\textwidth]{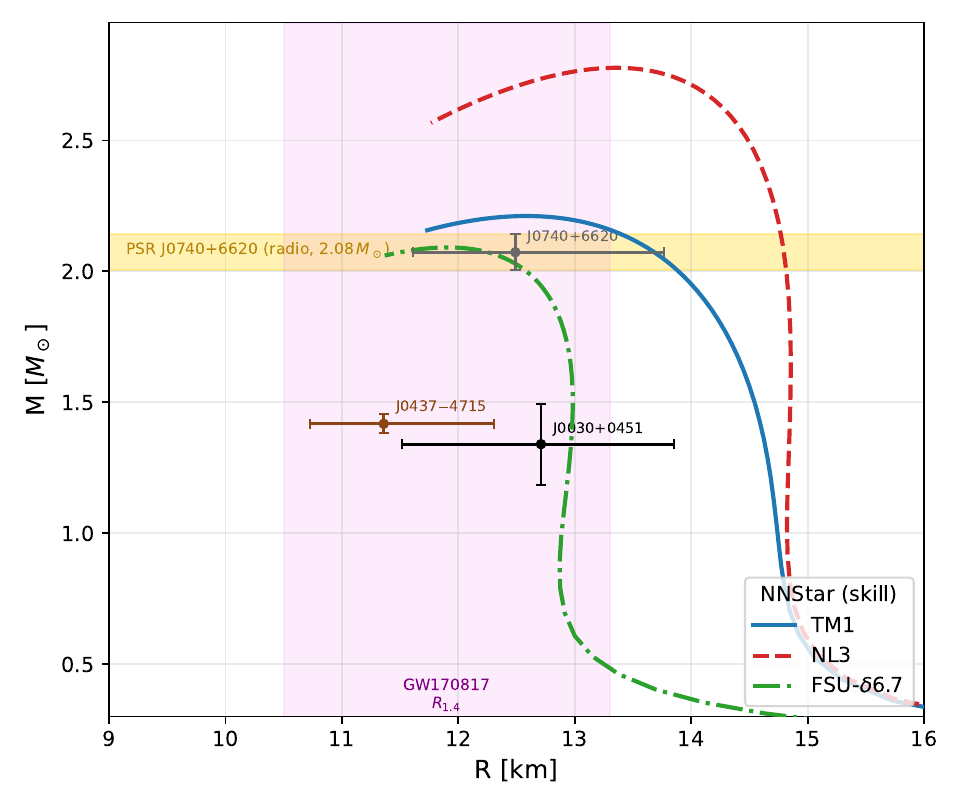}
\caption{Mass--radius curves for TM1, NL3, and FSU-$\delta$6.7 generated by the \textsc{NNStar} skill (couplings from Ref.~\cite{Ma:2026kun}), overlaid on current astrophysical constraints: the radio mass of PSR~J0740+6620 (gold band)~\cite{Fonseca:2021wxt}; the NICER mass--radius measurements of PSR~J0030+0451~\cite{Vinciguerra:2023qxq}, PSR~J0740+6620~\cite{Salmi:2024aum}, and PSR~J0437$-$4715~\cite{Choudhury:2024xbk} ($1\sigma$ error bars); and the GW170817 constraint on $R_{1.4}$ (violet band)~\cite{LIGOScientific:2018cki}. The stiff NL3 and TM1 forces yield large radii, while the $\delta$-meson model FSU-$\delta$6.7 is softer and lies closer to the data, reproducing the published behavior of these models. }
\label{fig:val-mr}
\end{figure}

The agreement in Table~\ref{tab:val-nmp} is at the rounding level for $n_0, E, E_{\rm sym}$, and $L$, while $K$ is systematically higher than the reference by $\sim18$~MeV ($\sim7\%$) for TM1 and NL3, but agrees to within $\sim1\%$ for FSU-$\delta$6.7. The dominant cause is the precision of the published inputs: the couplings in Table~II of Ref.~\cite{Ma:2026kun} are quoted to only three significant figures, and $K=9\,dP/dn$ is a second derivative of the energy per particle, making it far more sensitive to the scalar self-couplings $g_{s3},g_{s4}$, and the mass $m_\sigma$ than to the lower-order quantities $n_0, E, E_{\rm sym}$, and $L$. A truncation of these inputs at the third digit therefore propagates into a few-percent shift in $K$, while leaving the other observables within rounding. This interpretation is confirmed by the GQHD2 model, for which the full-precision couplings are available and the agent recovers $K=214$~MeV against the published value of $214$~MeV. Sub-percent residual differences arise from the uniform choice of meson masses ($m_\omega=783, m_\rho=770, m_{a_0}=980$~MeV), the fixed crust--core matching at $n_b=0.08~\mathrm{fm}^{-3}$, and the numerical determination of the saturation point. None of these affect the conclusion that the skill reproduces the published results; tightening $K$ for TM1 and NL3 would only require the original full-precision couplings.

\subsection{\texorpdfstring{Extending and optimizing a model: a $\sigma^6$-augmented TM1}{Extending and optimizing a model: a sigma6-augmented TM1}}
\label{sec:sigma6}

A more demanding test of the skill is not merely to reproduce a published model, but to extend its Lagrangian and re-optimize the new theory against data---the workflow a practitioner actually follows when proposing a model. We use the agent to add a sextic scalar self-interaction, $\mathcal{L}\supset-\tfrac{1}{6}g_{s6}\,\sigma^6$, to TM1 and to find the optimal couplings of the extended model.

The physics motivation is well established. The cubic and quartic scalar self-interactions of Boguta and Bodmer~\cite{Boguta:1977xi} are precisely what reduce the Walecka incompressibility from $K\sim550$~MeV to the empirical value of $K\sim230$~MeV; higher-order scalar terms continue this logic at supra-saturation density.
M\"uller and Serot~\cite{Mueller:1996pia} showed that models sharing the same saturation properties can differ in the NS maximum mass by more than $1\,M_\odot$ once the high-order scalar sector is varied. Maslov {\it et al}~\cite{Maslov:2015wba} demonstrated that a scalar potential rising steeply just above $n_0$ halts the decrease of the effective mass and stiffens the EoS at high density---raising $M_{\rm max}$---while leaving the saturation point untouched. Motohiro {\it et al}~\cite{Motohiro:2015taa} showed that a six-point interaction of
the $\sigma$ meson in the parity-doublet model reasonably reproduces the properties of normal NM. A $\sigma^6$ term is thus a natural high-density handle, and adding it to TM1 exercises both the symbolic (Lagrangian-to-EOM) and the optimization capabilities of the skill.

Optimization protocol.---The agent builds the augmented model by appending the single $-\tfrac{1}{6}g_{s6}\sigma^6$ term to the TM1 definition, then minimizes the objective
\begin{equation}
\chi^2(\theta)=\sum_i\left(\frac{R_i(\theta)-D_i}{\sigma_i}\right)^2 ,
\label{eq:chi2}
\end{equation}
where $R_i(\theta)$ is the value of observable $i$ computed by the skill's forward pipeline, and $(D_i,\sigma_i)$ are the empirical central value and $1\sigma$ width.
This is identical, up to an additive constant, to the BJA log-likelihood of Eq.~\eqref{eq:bjascore}: with the normalized residual of App.~\ref{app:BJA}, the rescaling factor $\Delta$ cancels in the exponent, so $\mathrm{BJA}(\theta)=-\tfrac{1}{2}\chi^2(\theta)+\text{const}$, and minimizing $\chi^2$ maximizes the BJA score.
The constraints $D_i\pm\sigma_i$ are the NM set of Ref.~\cite{Ma:2026kun}---$n_0=0.16\pm0.05$~fm$^{-3}$, $E=-16.0\pm1.0, K=230\pm30, E_{\rm sym}=30.9\pm1.9$, $L=52.5\pm17.5$~MeV, and $m_\sigma=600\pm200$~MeV---together with the astrophysical bounds $M_{\rm max}\geq2.08\,M_\odot$ (PSR~J0740+6620~\cite{Fonseca:2021wxt}, imposed as a one-sided penalty) and $R_{1.4}=12.7\pm1.1$~km (NICER PSR~J0030+0451~\cite{Vinciguerra:2023qxq}).
Starting from the TM1 couplings (with $g_{s6}=0$), all seven couplings $\{g_\sigma,g_\omega,g_\rho,g_{s3},g_{s4},c_3,g_{s6}\}$ and the scalar mass $m_\sigma$ are freed; the saturation observables are evaluated and the NS sector through the TOV solver, and the minimization is carried out by a gradient-based search refined over the high-density coupling $(g_{s6})$.

\begin{figure}[t]
\centering
\includegraphics[width=0.48\textwidth]{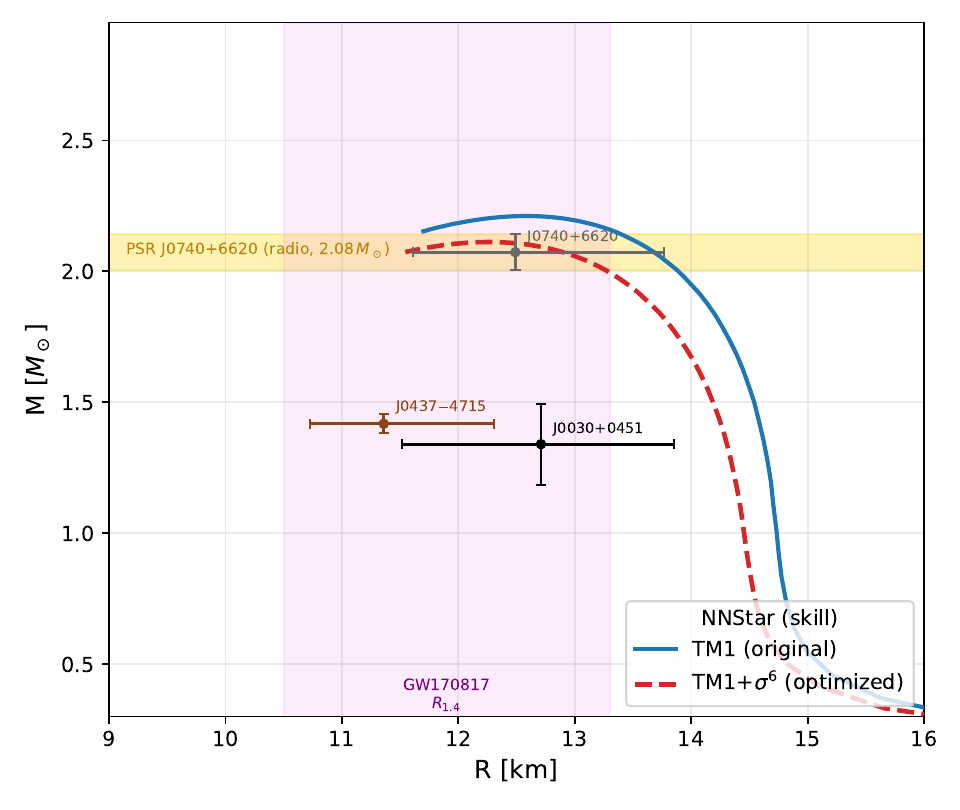}
\caption{Mass--radius curves of the original TM1 (solid) and the agent-optimized $\sigma^6$-augmented TM1 (dashed), with the same astrophysical constraints as in Fig.~\ref{fig:val-mr}. Adding the sextic scalar term and re-optimizing softens the equation of state slightly, shrinking the radius while keeping $M_{\rm max}>2\,M_\odot$.}
\label{fig:tm1-sigma6}
\end{figure}

Result.---The agent drives the total $\chi^2$ from $25.5$ for the original TM1 to $7.2$ for the optimized $\sigma^6$ model, with optimal couplings $m_\sigma=492.4, g_\sigma=-9.93, g_\omega=12.86, g_\rho=3.89, g_{s3}=1743, g_{s4}=-0.62, c_3=107.6$~(all in their natural units), and $g_{s6}=-2.9\times10^{-4}~\mathrm{MeV}^{-2}$. Table~\ref{tab:sigma6} compares the saturation and NS observables before and after the optimization, and Fig.~\ref{fig:tm1-sigma6} shows the corresponding M-R curves. The fit brings the incompressibility ($K:275\!\to\!231$~MeV) and the symmetry energy ($E_{\rm sym}:36.6\!\to\!30.1$~MeV) onto their empirical targets, and reduces the over-stiff slope ($L:110\!\to\!90$~MeV), all while preserving a maximum mass above $2\,M_\odot$ ($M_{\rm max}=2.11\,M_\odot$). The optimal $g_{s6}$ is small and negative: with the saturation sector already well constrained, the sextic term acts as a fine high-density adjustment that allows the agent meet the incompressibility and maximum-mass requirements simultaneously, rather than as a dominant new force. The residual tension lies in the radius, $R_{1.4}\simeq14.3$~km, which remains larger than the NICER value: $R_{1.4}$ is governed mainly by the symmetry energy below saturation and cannot be brought down by a purely isoscalar $\sigma^6$ term, an honest limitation that a richer isovector sector (as in the full GQHD model) is needed to address. The exercise nonetheless demonstrates the intended workflow end-to-end: the agent extends a published Lagrangian, derives the new field equations, and optimizes the enlarged coupling space against the full data set without human intervention.

\begin{table}[t]

\caption{\label{tab:sigma6}Saturation and neutron-star observables for the original TM1 and for the $\sigma^6$-augmented TM1 optimized by the agent against the empirical constraints (Eq.~\eqref{eq:chi2}). The last row is the total $\chi^2$. $M_{\rm max}$ enters as a one-sided bound ($\geq2.08\,M_\odot$).}
\begin{ruledtabular}
\begin{tabular}{lccc}
Observable & Empirical & TM1 & TM1$+\sigma^6$ \\
\hline
$n_0$ [fm$^{-3}$]        & $0.16\pm0.05$   & $0.145$ & $0.141$ \\
$E(n_0)$ [MeV]          & $-16.0\pm1.0$   & $-15.9$ & $-16.1$ \\
$K(n_0)$ [MeV]          & $230\pm30$      & $275$   & $231$ \\
$E_{\rm sym}(n_0)$ [MeV] & $30.9\pm1.9$   & $36.6$  & $30.1$ \\
$L(n_0)$ [MeV]          & $52.5\pm17.5$   & $110$   & $90$ \\
$m_\sigma$ [MeV]        & $600\pm200$     & $511$   & $492$ \\
$M_{\rm max}$ [$M_\odot$] & $\geq2.08$    & $2.21$  & $2.11$ \\
$R_{1.4}$ [km]          & $12.7\pm1.1$    & $14.6$  & $14.3$ \\
\hline
$\chi^2$                & --              & $25.5$  & $7.2$ \\
\end{tabular}
\end{ruledtabular}
\end{table}

\section{Benchmarks}
\label{sec:benchmark}

A central question for any scientific agent is how much its reliability depends on grounded use of the skill as opposed to the parametric knowledge of the underlying LLM~\cite{Zheng:2025survey}. A key advantage of the present setting is that, unlike open-ended literature or reasoning benchmarks~\cite{Ma:2024sciagent,Wang:2025paperarena}, dense-matter physics admits a deterministic ground truth: the correct value of any NM or NS observable for a given Lagrangian is fixed by the field equations and computed by the tool stack itself. This allows us to score the agent against exact references rather than human judgment-an outcome-based protocol in the spirit of recent agentic tool-use benchmarks~\cite{Lei:2025mcpverse}.

\subsection{Task suite and metrics}
\label{sec:bench-design}

Performance is measured along three orthogonal axes, following the decomposition used in function-calling and agentic-tool evaluations~\cite{Lei:2025mcpverse,Ma:2024sciagent}: (a) task success rate---the fraction of tasks whose final answer matches the reference within tolerance; (b) tool-call validity---whether the agent selects the correct tool and fills its arguments with a schema-correct call, separating intent understanding from format alignment; and (c) workflow efficiency---the number of tool calls and redundant or failed invocations relative to the optimal plan. Separating (a) from (b)--(c) is essential: a model may reach the right answer through an inefficient path, or issue valid calls yet misinterpret the result. In the present evaluation, we instantiate this protocol on the six observables of Sec.~\ref{sec:model}---the binding energy, pressure, incompressibility, and symmetry energy at $n_0$, together with the NS mass and radius---computed from a fixed, randomly sampled Lagrangian. For each LLM, we draw three independent tasks and solve each in both modes (skill-equipped and skill-free) under identical conditions, with cross-session memory disabled and one fresh context per task; the agent must return the six values as a numerical vector, which is scored against the deterministic reference produced by the validated tool stack. For every run, we record the per-observable accuracy together with the token usage and wall-clock time (Tables~\ref{tab:bench-llm} and~\ref{tab:bench-detail}). This is a deliberately compact, proof-of-concept evaluation rather than an exhaustive one.

\subsection{With and without the skill}
\label{sec:bench-tools}

We run each LLM on the suite in two modes: (i) skill-equipped, in which the agent has loaded the \textsc{NNStar} skill and executes its bundled, validated physics code; and (ii) skill-free, in which the same LLM receives the identical task but without the skill and must answer from its own reasoning and ad hoc code alone.
The comparison isolates the value added by the skill and quantifies the error incurred when the model is forced to ``recall'' or improvise rather than run grounded, validated code. 
To keep the comparison clean, the platform's cross-session memory is disabled and each task is solved in a fresh context, so the only difference between the two modes is the presence of the skill.
Table~\ref{tab:bench-llm} reports this comparison for six contemporary LLMs, while App.~\ref{app:bench} provides the complete per-run results in Table~\ref{tab:bench-detail}.
With the skill, every model that engages the bundled code reproduces the four NM quantities essentially exactly and the NS mass and radius \skl{to within about a percent (sub-percent apart from a single DeepSeek run whose TOV step errored out, Table~\ref{tab:bench-detail})}. Without it, the same models either reconstruct the pipeline only approximately or fail outright, with characteristic errors---most strikingly, a negative incompressibility $K<0$ from the saturation-only formula that the skill explicitly forbids (e.g.\ Claude and DeepSeek in Table~\ref{tab:bench-detail}). The skill thus turns a knowledge-dependent, unreliable computation into a deterministic one.

\subsection{Across different LLMs}
\label{sec:bench-llm}

Because the skill is model-agnostic, it can be loaded by different LLMs. We benchmark a representative set---DeepSeek, Qwen, Kimi, GPT, Claude, and Gemini---on the identical suite and metrics.
Table~\ref{tab:bench-llm} summarizes this comparison.
The skill-equipped accuracy is uniformly improved across all models, except that Kimi and one set of Qwen runs fail to engage the skill and instead return a hallucinated answer or nothing available.
Without the skill, the models' performance is poor and inconsistent: sometimes nothing is returned, and sometimes the agent misapplies the physics.
Accordingly, the benchmark reported here is intended as an initial demonstration of the with/without-skill and cross-model effects rather than a final score; a more rigorous and larger-scale benchmarking scheme---with a broader task distribution, more samples per model, and a formally defined scoring metric---will be presented in a forthcoming companion paper focused on the agent algorithm.

\begin{table*}[t]

\caption{\label{tab:bench-llm}Cross-LLM benchmark: three random tasks per model, each solved with and without the \textsc{NNStar} skill (paired, identical tasks). ``Valid'' counts runs returning a well-formed six-vector; the error columns are \skl{means} over valid runs of the relative error (\%) on the nuclear-matter quantities (worst of $E/A,P,K,E_{\rm sym}$), the mass $M$, and radius $R$ versus the deterministic reference. ``Tokens'' (thousands) and ``Time'' (s) are means and measure efficiency.}
\begin{ruledtabular}
\begin{tabular}{llcccccc}
Model & Mode & Valid & matter (\%) & $M$ (\%) & $R$ (\%) & Tokens (k) & Time (s) \\
\hline
GPT-5.5 & $+$skill & 3/3 & 0.0 & 0.0 & 0.0 & 65 & 109 \\
 & $-$skill & 3/3 & 0.0 & \skl{7.7} & \skl{16.0} & 52 & 204 \\
\hline
Claude-Opus-4.8 & $+$skill & 3/3 & 0.0 & \skl{0.3} & \skl{0.8} & 77 & 103 \\
 & $-$skill & 3/3 & \skl{186.5} & \skl{34.8} & \skl{17.0} & 116 & 352 \\
\hline
Gemini-3.1-Pro & $+$skill & 3/3 & 0.0 & \skl{0.1} & \skl{0.3} & 1218 & 386 \\
 & $-$skill & 2/3 & 52.4 & 1.0 & 1.0 & 1341 & 364 \\
\hline
Qwen3.7-Max & $+$skill & 2/3 & 0.0 & 0.0 & 0.0 & 112 & 161 \\
 & $-$skill & 0/3 & -- & -- & -- & -- & 154 \\
\hline
DeepSeek-V4-Pro & $+$skill & 3/3 & 0.0 & \skl{2.4} & \skl{1.7} & 63 & 394 \\
 & $-$skill & 2/3 & 2214.3 & 100.0 & \skl{$\sim\!10^{10}$} & 55 & 606 \\
\hline
Kimi-K2.6 & $+$skill & 2/3 & 101.1 & 79.9 & 75.0 & 75 & 605 \\
 & $-$skill & 2/3 & 1334.4 & 100.0 & 85.7 & 56 & 609 \\
\end{tabular}
\end{ruledtabular}
\end{table*}

\section{Conclusion and outlook}
\label{sec:conclusion}

We have presented \textsc{NNStar}, a portable \emph{skill} that turns a general LLM agent into an end-to-end solver which automates the labor- and time-intensive workflow of confronting a hadronic effective model with the full range of NM and NS data.
By separating an LLM reasoning layer from a deterministic, layered physics stack---packaged as a platform-agnostic skill that the agent loads and executes---the agent builds a model from its Lagrangian, derives and solves the mean-field EOMs, computes the saturation properties, constructs the $\beta$-equilibrium EoS with a realistic crust, integrates the TOV equations, and assesses the result through a BJA---all driven by natural-language instructions.
The agent can read a model, reproduce its predictions, and refine its couplings.
It has been validated against the Walecka-type models TM1, NL3, and FSU-$\delta$6.7, reproducing their published saturation properties and M-R curves.
Going beyond reproduction, we used the agent to an extended model---adding a sextic scalar self-interaction $\sigma^6$ to TM1 and re-optimizing the enlarged coupling space against the full NM and NS constraint set---and we benchmarked several contemporary LLMs driving the identical skill, which shows that grounding in the bundled code turns an otherwise knowledge-dependent and unreliable computation into a deterministic one, largely independent of the underlying model.

Several directions remain to be investigated in the future.
One possible physics extension is to implement chiral effective models within the same skill, making it possible to study how chiral symmetry constrains the dense-matter EoS through the same automated workflow~\cite{Ma:2025llw,Xiong:2025jxq}.
On the agent side, the benchmarks of Sec.~\ref{sec:benchmark} should be completed and extended, and the skill's tool catalog can be broadened to include additional observables (e.g., finite-temperature EoS and transport coefficients) and additional classes of models beyond the nucleon-only description---such as the inclusion of quark matter in the cores of massive NSs and strangeness through hyperons. Because the capability is delivered as a skill, such extensions immediately benefit any agent that loads it. More broadly, by packaging a complete, auditable physics stack as a shareable skill that any capable LLM can load and drive, this work points to a new mode of research for dense-matter physics: the tedious forward-and-inference pipeline is delegated to a grounded AI agent, while the physicist focuses on proposing models and posing questions. We expect that grounding LLM reasoning in verifiable, shareable physics skills, as done here, will become an increasingly important methodology for dense-matter phenomenology and, more generally, for confronting effective models with heterogeneous data.

\acknowledgments

The work of Y. M. is supported by Jiangsu Funding Program for Excellent Postdoctoral Talent under Grant Number 2025ZB516. Y.~L. M. is supported in part by the National Science Foundation of China (NSFC) under Grant No. 12547104, the National Key R\&D Program of China under Grant No. 2021YFC2202900 and Gusu Talent Innovation Program under Grant No. ZXL2024363.

\section*{Data availability}
All process-related data and code---the benchmark task suite, the deterministic ground-truth values, the \textsc{NNStar} skill, and the agent-orchestration and scoring scripts used to produce the results of Sec.~\ref{sec:benchmark}---are publicly available on GitHub at \url{https://github.com/AaronMahn/NM-rmf-benchmark-scheme}.


\begin{widetext}

\appendix

\section{Bayesian joint analysis}
\label{app:BJA}

Based on Bayes' theorem~\eqref{eq:bayes} and discussion in the main text, the quantity of interest is the posterior $p(\theta\mid D)$.
Using Eq.~\eqref{eq:like}, the posterior is then
\begin{equation}
p(\theta\mid D)\;\propto\;\prod_{i=1}^{n} p(D_i\mid\theta)\,p(\theta).
\end{equation}
We assume a normalized likelihood for each observable and uniform priors for simplicity (the extension to other distributions is straightforward) within the allowed parameter ranges. To ensure comparable contributions from observables with vastly different magnitudes and uncertainties, all data are mapped to a normalized residual space in which the experimental central values are shifted to zero and the uncertainties rescaled to a common effective value $\Delta=0.5$. Explicitly, for each observable, we define the dimensionless normalized residual $\widetilde{M}_i(\theta)$ which measures the difference between the model prediction and the experimental central value in units of the experimental uncertainty, rescaled by $\Delta$:
\begin{equation}
\widetilde{M}_i(\theta)=\Delta\,\frac{R_i(\theta)-D_i^{\rm exp}}{\sigma_i^{\rm exp}},
\end{equation}
where $D_i^{\rm exp}$ and $\sigma_i^{\rm exp}$ are, respectively, the experimental central value and uncertainty of the $i$-th observable and $R_i(\theta)$ is the model prediction. The corresponding likelihood is
\begin{equation}
p(D_i\mid\theta)=\frac{1}{\sqrt{2\pi}\,\Delta}\,
\exp\!\left(-\frac{\widetilde{M}_i^2(\theta)}{2\Delta^2}\right).
\end{equation}
Because the same $\Delta$ multiplies every residual, it cancels between $\widetilde{M}_i^2$ and the $\Delta^2$ in the exponent; each observable therefore enters through its own inverse-variance weight $1/(\sigma_i^{\rm exp})^2$. As a result, quantities with larger absolute magnitude do not dominate the inference, while $\Delta$ only sets a common overall normalization of the likelihood. Finally, to avoid numerical overflow from the product of likelihood terms, the BJA score used to rank models is defined as
\begin{equation}
\mathrm{BJA}=\ln\!\left(\prod_{i=1}^{n} p(D_i\mid\theta)\,p(\theta)\right),
\label{eq:bjascore}
\end{equation}
with higher BJA indicating better overall agreement between the model predictions and the empirical data. In the agent, the M-R contribution to Eq.~\eqref{eq:bjascore} is estimated by taking the maximum value of the model-predicted M-R line for a given constraint.

For uniform priors $p(\theta)=\text{const}$ within the allowed ranges, substituting the likelihood into Eq.~\eqref{eq:bjascore} and using the cancellation of $\Delta$ noted above gives the explicit connection to the least-squares objective of Eq.~\eqref{eq:chi2},
\begin{equation}
\mathrm{BJA}(\theta)=-\tfrac{1}{2}\,\chi^2(\theta)+C,
\qquad
C=-n\!\left(\tfrac{1}{2}\ln 2\pi+\ln\Delta\right)+\ln p(\theta),
\label{eq:bja-chi2}
\end{equation}
where $\chi^2(\theta)=\sum_{i=1}^{n}\big[(R_i(\theta)-D_i)/\sigma_i\big]^2$ and the constant $C$ collects the per-observable likelihood normalization and the (constant) prior. Since $C$ is independent of $\theta$, maximizing the BJA score is exactly equivalent to minimizing $\chi^2$.
The only term not of this form is the one-sided maximum-mass bound $M_{\rm max}\geq2.08\,M_\odot$ used in Sec.~\ref{sec:sigma6}, which is imposed as a penalty active only when the bound is violated and therefore contributes to $C$ wherever the bound is satisfied.

\clearpage

\section{Per-run benchmark data}
\label{app:bench}

To make the aggregate results in Table~\ref{tab:bench-llm} auditable, Table~\ref{tab:bench-detail} reports the ground-truth and agent-computed observables, token usage, and wall-clock time for every run. These per-run data reveal failures that are not visible in the averaged results and distinguish invalid outputs from errors in the NM calculation or TOV step.
{
\begingroup
\renewcommand{\arraystretch}{1.0}%
\footnotesize
\setlength{\tabcolsep}{5pt}%
\newcommand{\rowsep}{\rule[-1.5mm]{0pt}{5.2mm}}
\begin{center}
\refstepcounter{table}\label{tab:bench-detail}
\edef\benchdetailtabnum{\Roman{table}}
\noindent\begin{minipage}{0.96\textwidth}
\footnotesize\textbf{TABLE~\benchdetailtabnum.} Ground truth (GT) vs.\ agent-computed values for every run.
``$+$skill''/``$-$skill'' are the two modes; ``inv.''\ marks a run with no valid
six-vector; ``--'' denotes not applicable. $n$ is the task index.
\end{minipage}
\vspace{0.6em}

\begin{longtable}{l c >{\rowsep}l cccc cc cc}
\hline
Model & $n$ & Mode & $E/A$ & $P$ & $K$ & $E_{\rm sym}$ & $M$ & $R$ & Tok & Time \\
 & & & [MeV] & [MeV\,fm$^{-3}$] & [MeV] & [MeV] & [$M_\odot$] & [km] & [$10^3$] & [s] \\
\hline
\endfirsthead
\multicolumn{11}{l}{\footnotesize\itshape TABLE~\benchdetailtabnum\ (continued)}\\
\hline
Model & $n$ & Mode & $E/A$ & $P$ & $K$ & $E_{\rm sym}$ & $M$ & $R$ & Tok & Time \\
 & & & [MeV] & [MeV\,fm$^{-3}$] & [MeV] & [MeV] & [$M_\odot$] & [km] & [$10^3$] & [s] \\
\hline
\endhead
\midrule
\multicolumn{11}{r}{\footnotesize\itshape continued on next page}\\
\endfoot
\hline
\endlastfoot
\multirow{9}{*}{GPT-5.5}
 & \multirow{3}{*}{8}  & GT      & -26.1 & -0.9 & 301 & 56.8 & 1.719 & 8.78  & --    & --  \\
 &                     & $+$skill& -26.1 & -0.9 & 301 & 56.8 & 1.719 & 8.78  & 53.8  & 74  \\
 &                     & $-$skill& -26.1 & -0.9 & 301 & 56.8 & 2.037 & 9.92  & 60.1  & 264 \\
\cline{2-11}
 & \multirow{3}{*}{32} & GT      & 45.5  & 5.7  & 511 & 16.1 & 1.529 & 9.23  & --    & --  \\
 &                     & $+$skill& 45.5  & 5.7  & 511 & 16.1 & 1.529 & 9.23  & 59.8  & 70  \\
 &                     & $-$skill& 45.5  & 5.7  & 511 & 16.1 & 1.496 & 10.93 & 17.2  & 141 \\
\cline{2-11}
 & \multirow{3}{*}{46} & GT      & 101.0 & 12.6 & 1072& 22.0 & 1.562 & 10.39 & --    & --  \\
 &                     & $+$skill& 101.0 & 12.6 & 1072& 22.0 & 1.562 & 10.39 & 81.0  & 183 \\
 &                     & $-$skill& 101.0 & 12.6 & 1072& 22.0 & 1.599 & 12.10 & 79.1  & 206 \\
\hline
\multirow{9}{*}{Claude-Opus-4.8}
 & \multirow{3}{*}{23} & GT      & 63.7  & 8.8  & 856  & 23.3 & 1.601 & 10.11 & --    & --  \\
 &                     & $+$skill& 63.7  & 8.8  & 856  & 23.3 & 1.600 & 10.05 & 80.7  & 78  \\
 &                     & $-$skill& 158.7 & 12.3 & -1477& 61.3 & 1.347 & 7.43  & 162.2 & 489 \\
\cline{2-11}
 & \multirow{3}{*}{94} & GT      & 125.6 & 19.5 & 2206 & 39.1 & 1.773 & 11.13 & --    & --  \\
 &                     & $+$skill& 125.6 & 19.5 & 2206 & 39.1 & 1.772 & 11.08 & 63.8  & 142 \\
 &                     & $-$skill& 125.6 & 19.5 & 16   & 39.1 & 1.771 & 11.05 & 98.0  & 307 \\
\cline{2-11}
 & \multirow{3}{*}{96} & GT      & 146.9 & 20.1 & 1941 & 62.3 & 1.873 & 14.30 & --    & --  \\
 &                     & $+$skill& 146.9 & 20.1 & 1941 & 62.3 & 1.861 & 14.12 & 86.1  & 90  \\
 &                     & $-$skill& 146.2 & 57.9 & -322 & 61.7 & 3.531 & 17.70 & 87.1  & 258 \\
\hline
\multirow{9}{*}{Gemini-3.1-Pro}
 & \multirow{3}{*}{11} & GT      & 104.8 & 13.3 & 1151 & 29.3 & 1.332 & 10.02 & --     & --  \\
 &                     & $+$skill& 104.8 & 13.3 & 1151 & 29.3 & 1.332 & 10.00 & 1647.6 & 560 \\
 &                     & $-$skill& 104.8 & 13.3 & 1151 & 29.3 & 1.319 & 9.83  & 1636.1 & 533 \\
\cline{2-11}
 & \multirow{3}{*}{45} & GT      & 72.7  & 8.9  & 752  & 27.0 & 1.343 & 9.31  & --     & --  \\
 &                     & $+$skill& 72.7  & 8.9  & 752  & 27.0 & 1.343 & 9.31  & 1015.6 & 274 \\
 &                     & $-$skill& \textit{inv.} & \textit{inv.} & \textit{inv.} & \textit{inv.} & \textit{inv.} & \textit{inv.} & --     & 108 \\
\cline{2-11}
 & \multirow{3}{*}{62} & GT      & 48.0  & 7.6  & 817  & 14.7 & 1.642 & 10.03 & --     & --  \\
 &                     & $+$skill& 48.0  & 7.6  & 817  & 14.7 & 1.646 & 10.08 & 991.7  & 324 \\
 &                     & $-$skill& 48.0  & 7.6  & -39  & 14.7 & 1.627 & 10.00 & 1044.9 & 453 \\
\hline
\multirow{9}{*}{Qwen3.7-Max}
 & \multirow{3}{*}{6}  & GT      & 118.2 & 18.1 & 2027 & 52.4 & 1.657 & 10.58 & --    & --  \\
 &                     & $+$skill& 118.2 & 18.1 & 2027 & 52.4 & 1.657 & 10.58 & 89.2  & 97  \\
 &                     & $-$skill& \textit{inv.} & \textit{inv.} & \textit{inv.} & \textit{inv.} & \textit{inv.} & \textit{inv.} & --    & 155 \\
\cline{2-11}
 & \multirow{3}{*}{8}  & GT      & -26.1 & -0.9 & 301  & 56.8 & 1.719 & 8.78  & --    & --  \\
 &                     & $+$skill& -26.1 & -0.9 & 301  & 56.8 & 1.719 & 8.78  & 135.1 & 197 \\
 &                     & $-$skill& \textit{inv.} & \textit{inv.} & \textit{inv.} & \textit{inv.} & \textit{inv.} & \textit{inv.} & --    & 154 \\
\cline{2-11}
 & \multirow{3}{*}{93} & GT      & -11.2 & -0.7 & 60   & 19.5 & 1.603 & 8.91  & --    & --  \\
 &                     & $+$skill& \textit{inv.} & \textit{inv.} & \textit{inv.} & \textit{inv.} & \textit{inv.} & \textit{inv.} & --    & 190 \\
 &                     & $-$skill& \textit{inv.} & \textit{inv.} & \textit{inv.} & \textit{inv.} & \textit{inv.} & \textit{inv.} & --    & 153 \\
\hline
\pagebreak
\multirow{9}{*}{DeepSeek-V4-Pro}
 & \multirow{3}{*}{57} & GT      & 50.5   & 8.1  & 876  & 19.6  & 1.657 & 10.62 & --   & --  \\
 &                     & $+$skill& 50.5   & 8.1  & 876  & 19.6  & 1.657 & 10.62 & 54.2 & 229 \\
 &                     & $-$skill& -305.9 & 84.4 & 4912 & 376.9 & 0.000 & $2.0{\times}10^{9}$ & 53.1 & 606 \\
\cline{2-11}
 & \multirow{3}{*}{66} & GT      & 111.2 & 14.5 & 1296 & 22.7 & 1.564 & 10.49 & --   & --  \\
 &                     & $+$skill& 111.2 & 14.5 & 1296 & 22.7 & 1.564 & 10.49 & 58.0 & 345 \\
 &                     & $-$skill& \textit{inv.} & \textit{inv.} & \textit{inv.} & \textit{inv.} & \textit{inv.} & \textit{inv.} & --   & 607 \\
\cline{2-11}
 & \multirow{3}{*}{71} & GT      & 34.7  & 4.5  & 416  & 24.0  & 1.361 & 8.52  & --   & --  \\
 &                     & $+$skill& 34.7  & 4.5  & 416  & 24.0  & 1.263 & 8.07  & 78.2 & 607 \\
 &                     & $-$skill& 939.0 & 0.0  & -0   & -0.0  & 0.000 & 0.00  & 57.4 & 607 \\
\hline
\multirow{9}{*}{Kimi-K2.6}
 & \multirow{3}{*}{16} & GT      & 181.8 & 23.8 & 2153 & 35.1 & 1.663 & 13.40 & --   & --  \\
 &                     & $+$skill& 21.8  & 2.3  & -45  & 11.8 & 0.667 & 6.70  & 98.4 & 605 \\
 &                     & $-$skill& \textit{inv.} & \textit{inv.} & \textit{inv.} & \textit{inv.} & \textit{inv.} & \textit{inv.} & --   & 618 \\
\cline{2-11}
 & \multirow{3}{*}{21} & GT      & 128.0  & 16.7 & 1504  & 30.1 & 1.513 & 11.68 & --   & --  \\
 &                     & $+$skill& \textit{inv.} & \textit{inv.} & \textit{inv.} & \textit{inv.} & \textit{inv.} & \textit{inv.} & --   & 605 \\
 &                     & $-$skill& -469.5 & 75.1 & 4226  & -0.0 & 0.000 & 20.00 & 51.1 & 605 \\
\cline{2-11}
 & \multirow{3}{*}{36} & GT      & 130.4  & 17.4 & 1612  & 21.1 & 1.409 & 11.21 & --   & --  \\
 &                     & $+$skill& 0.0    & 0.0  & -0    & 0.1  & 0.000 & 0.00  & 52.1 & 604 \\
 &                     & $-$skill& -469.5 & 0.4  & 37105 & 0.1  & 0.000 & 0.00  & 60.4 & 604 \\
\end{longtable}
\end{center}
\endgroup
}

\clearpage

\end{widetext}

\bibliography{AIAgentRef}

\end{document}